# Search for Short Duration Bursts of TeV Gamma Rays with the Milagrito Telescope

Gus Sinnis[1] for the Milagro Collaboration

[1] *Los Alamos National Laboratory, Los Alamos, NM 87545, USA*

### Abstract
The Milagrito water Cherenkov telescope operated for over a year (2/97-5/98). The most probable gamma-ray energy was ~1 TeV and the trigger rate was as high as 400 Hz. Milagrito has opened a new window on the TeV Universe. We have developed an efficient technique for searching the entire sky for short duration bursts of TeV photons. Such bursts may result from "traditional" gamma-ray bursts that were not in the field-of-view of any other instruments, the evaporation of primordial black holes, or some as yet undiscovered phenomenon. We have begun to search the Milagrito data set for bursts of duration 10 seconds. Here we will present the technique and the expected results. Final results will be presented at the conference.

## 1  Introduction:

The Milagrito detector is described in detail elsewhere (Westerhoff 1999). The detector operated between February 1997 and May 1998. The event rate varied from 200 Hz to 400 Hz, depending upon the water depth above the photomultiplier tubes (PMTs) and the accumulated rainfall and snowmelt on top of the detector. The energy threshold of Milagrito was below 1 TeV and the most probable energy for gamma rays from zenith was ~1 TeV. After a software threshold was applied to the data (we required at least 40 photomultiplier tubes be used in the angle fit), 64% of the data survived and the angular resolution of these events was ~0.7° (Wascko 1999). The data set consisted of nearly 6 billion events. Because searching this data set for short duration bursts is computationally intensive; we have developed an algorithm that is relatively efficient. The entire analysis can be run in 10-20 days on a 7-node PC farm.

While there are several reasons to search for short duration bursts (TeV counterparts to gamma-ray bursts, the final stages of black hole evaporation) the most compelling reason may be the discovery potential of a new phenomenon. To our knowledge this is the first such search in this energy regime. In this paper we describe the technique, show results for a subset of the data and calculate the expected sensitivity of the search.

### Analysis Technique:

The analysis is a straightforward binned analysis. If one uses square bins, the optimal bin size is 2.8 times the angular resolution of the instrument (Alexandreas, et al., 1993). Given the angular resolution of 0.7 degrees we have used a bin with a 2.0 degree span in declination, and 2.0/cos($\delta$) in right ascension. To find the expected background in a given (right ascension/declination) bin we integrate the measured detector efficiency (in local coordinates) over the exposure of that bin. This number is then compared to the actual number of events that fell in the bin. A detailed description of the method follows. For simplicity we have to chosen to use hour angle (HA) and declination ($\delta$) as the local coordinate system.

**2.0 Background Estimation:** The number of events expected in a given time interval from a given direction in the sky is:

$$N_{\exp}(\Delta HA, \Delta \delta) = \iiint \varepsilon[HA(RA,t), \delta] R(t) d(HA) d\delta dt$$

where $\varepsilon$ is the efficiency of the detector as a function of the local coordinate system and R(t) is the overall event rate of the detector. The integral is over the angular bin and the time interval in question. The above equation is only correct if the detector efficiency is constant over the time interval used to estimate the background. The efficiency function ε(HA,δ) is obtained from the data in the following manner. Data is collected into maps (2-dimensional arrays) of 0.2 x 0.2 degree bins (of HA and δ) over two hour intervals. Each map contains the number of events collected in each small bin divided by the total number of events collected over the two-hour interval. This map is therefore a representation of the efficiency of the detector as a function of the local coordinate hour angle and declination. A two-hour interval was chosen to minimize systematic effects caused by any changes in the efficiency function and to obtain sufficient statistics to parameterize the background. To accommodate changes in the detector the integration interval could be shorter than two hours if the detector configuration changed. The background map for each interval was saved to disk.

To perform the integral given above we construct two additional maps of the sky, one for the background and one for the observed sky. Although these maps are in 0.2 x 0.2-degree bins, like the efficiency maps, these maps are in "sky" coordinates (right ascension and declination). The observed sky map is constructed by incrementing the appropriate bin for each observed event. The background map is constructed by incrementing every bin by its efficiency (the fraction of events in the contemporaneous 2-hour interval that fell within this bin). Since a simple movement in time relates the right ascension and the hour angle, the updating of the background array is accomplished by rotating the efficiency map to the correct local apparent sidereal time. To improve the performance of the search the background map is updated once every 10 seconds (by $N_{events}$ x ε(HA,δ)). (Note that the sky moves by 0.04 degrees in 10 seconds.)

**2.1 Search for an Excess:** Since we do not know the start time or the direction of a possible burst we must oversample the sky in both time and space. We use two time bins (each of ten seconds duration), shifted by 5 seconds, and 4 different grids of angular bins. If the first angular grid is centered on (RA,δ), the remaining grids are centered on (RA+1°/cos(δ),δ), (RA,δ+1°), and (RA+1°/cos(δ),δ+1°). Every 5 seconds a 10 second interval completes and all 4 angular grids are searched for an excess. In practice the 4 grids are formed "on-the-fly" by performing the appropriate sums over the 0.2 x 0.2 degree maps. These sums yield the number of observed events ($N_{obs}$) and the number of expected events ($N_{exp}$) in every 2.0 x 2.0/cos(δ) bin. The Poisson probability is calculated and the result is stored in a histogram. If the interval has a Poisson probability less then $10^{-8}$ the start time, right ascension, declination, $N_{exp}$, and $N_{obs}$ are also saved.

# 3 Search Results:

The probability distribution from one day of data is shown in Figure 1. Note that the figure does not contain completely independent entries. The results from each of the 4 angular grids are summed into a single histogram, as are the results from the two offset time bins. The spikes in the distributions are caused by the quantization of the observed and expected number of events. The slope of the distribution is consistent with expectations. No significant burst has been observed in this subset of the data. Figures 2 and 3 show the distribution of the expected number of events and the observed number of events from the same day.

**3.1 Flux Upper Limits:** If the search completes and no significant burst is found several upper limits may be given. Given the maximum number of observed events over the entire observation period (~1 year), one can derive an absolute upper limit for the entire time period. One can also derive a "typical" upper limit, based on the typical number of events observed in a bin. We report an expected absolute upper limit as a function of zenith angle (since the sensitivity of the detector changes with the zenith angle). This should give an indication of the sensitivity of our search. "Typical" upper limits are roughly 1/4 of the strict upper limit.

**3.2 Strict Upper Limit:** For the subset of the data searched the maximum number of observed events in any 10-second interval was 15 with a background of 2. Therefore, our 90% confidence level upper limit for the number of source events in any bin is 19.3. Three corrections must be applied to this number before it can be converted into a flux. First, no more than 48% of any source events should be contained in the signal bin. For a Gaussian response the fraction is 72%, however the angular resolution function of Milagrito has a significant non-Gaussian tail. In addition, in this analysis the sky was binned into 4 overlapping grids, thus there is an efficiency associated with the location of a source falling randomly on the sky. In practice these two effects must be accounted for simultaneously. For the worst case source location the combined correction is 2.4. Finally, two time windows shifted by 5 seconds are used. Thus a 10-second burst starting 2.5 seconds into a window will only have 75% of its events within any time window. Thus, in this scenario if no significant burst were observed the 90% C.L. upper limit on the number of excess events within any 10-second window would be 61.7 = 19.3x2.4x1.3. To convert this to an upper limit on the flux of gamma rays we must convolute the effective area of Milagrito with an assumed source spectrum, $I_0(E)^{-\alpha}$. If we assume, $\alpha=2$, we may set an upper limit on $I_0$. The effective area vs. energy for several ranges of zenith angle is shown in these proceedings (Figure 1 from Westerhoff 1999). The resulting expected upper limits to the flux are tabulated in Table 1.

| Zenith Angle Range (degrees) | Median Energy (TeV) | Expected Flux Upper Limit ($\gamma$'s cm$^{-2}$ sec$^{-1}$ TeV$^{-1}$) ($I_0$) |
|---|---|---|
| 0-15 | 7.5 | 7.8 x 10$^{-8}$ |
| 15-30 | 9.7 | 1.29 x 10$^{-7}$ |
| 30-45 | 13.2 | 3.76 x 10$^{-7}$ |

Table 1: Expected 90% C.L. upper limits to the flux of gamma rays for 10-second bursts viewed by Milagrito.

## 4 Future Work

The software cuts and the size of the angular bin used for this subset of the data has not been optimized for this particular search. Since the expected number of background events is small, the search bin should be larger (Alexandreas 1993). In addition since the source location within the bin is unknown one would expect the optimal bin to be larger yet. Monte Carlo work is in progress to optimize the search technique for this data set, and the response of the Milagrito detector.

## 5 Conclusions

We have begun to search the Milagrito data set for 10-second bursts from any direction of the sky. So far we have failed to detect any significant bursts of this duration. We have given an indication of the sensitivity of the method by reporting expected flux upper limits in the absence of any detected bursts. Final results will be reported at the conference.

## Acknowledgements

This research was supported in part by the National Science Foundation, the U. S. Department of Energy Office of High Energy Physics, the U. S. Department of Energy Office of Nuclear Physics, Los Alamos National Laboratory, the University of California, the California Space Institute, and the Institute of Geophysics and Planetary Physics. We would also like to acknowledge the hard work and dedication of Scott Delay, Michael Schneider, and Neil Thompson, without whom Milagro would still be a dream.

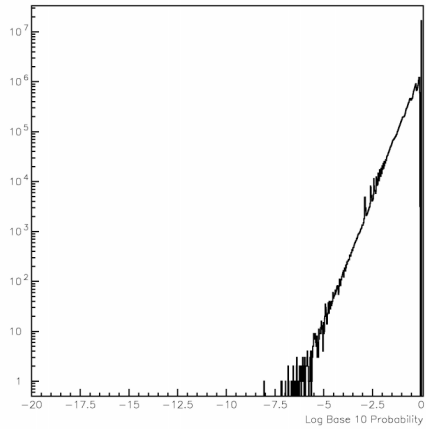

**Figure 1: Sample probability distribution from 10-second burst search.**

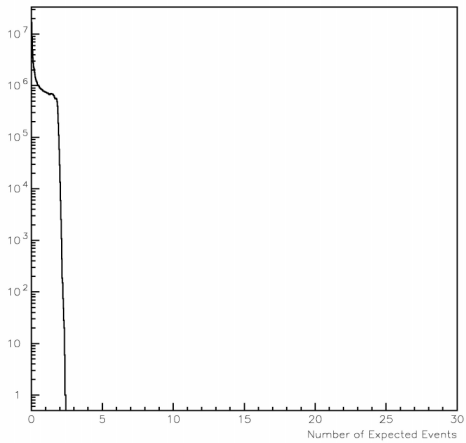

**Figure 2: Sample distribution of expected number of events from 10-second burst search.**

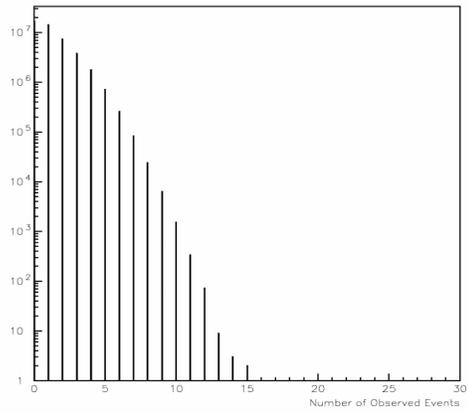

**Figure 3: Sample distribution of observed number of events from 10-second burst search.**